# Economic Determinants of Happiness:
# Evidence from the US General Social Survey

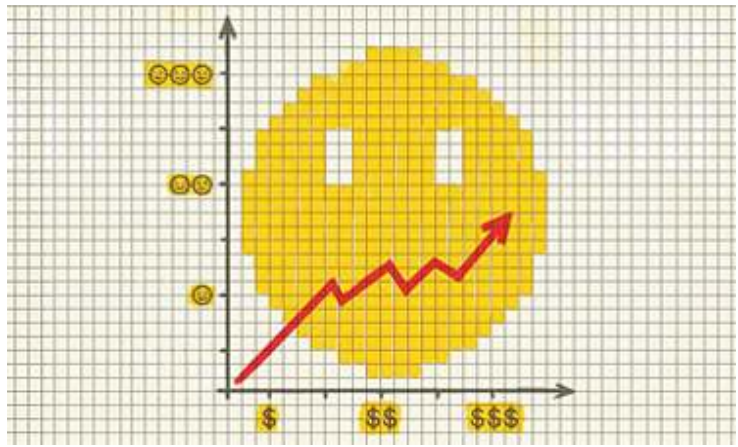

**Teng Guo (997106803)**
**Lingyi Hu(996703866)**

# Introduction

As the ancient philosopher Marcus Aurelius Antonius concluded after considerable thought into the concept of happiness "Very little is needed to make a happy life". In today's materialistic society it would seem that such insights are erroneous at its core conception. As people of our generation constantly travel on the high way to wealth in the pursuit of happiness; is their ultimate destination the one that they intended? More importantly will the nation as a whole achieve greater happiness as a result of increased national wealth and more favourable economic conditions? Though intuition and public opinion suggests that more income is always preferable to less, there has been considerable debate in this field when elevated to the national level. Many scholars have recently begun to dispute the assumed link between individual wellbeing and economic conditions and the extent to which the latter matters (Easterlin, 1995; Stevenson and Wolfers 2008; Tella and MacCulloch 2008). This dilemma is empirically demonstrated in the Latin America Public Opinion Project (LAPOP, 2011), which surveyed North and Latin America in terms of perceived life satisfaction. Higher measures found in the less developed countries of Brazil, Costa Rica, and Panama than in North America pose an intriguing quandary to traditional economic theory. In light of this predicament this paper aims to construct a sensible measure of the national happiness level for the United States on a year by year basis; and regress this against indicators of the national economy to provide insight into this puzzling enigma between national happiness and economic forces

Though personal well being appears to be in the domain of sociology or psychology, its implications for the economy are valuable and manifold. The government exists for the purpose of serving the people, and in the US if the "pursuit of happiness" is a right protected by the Declaration of Independence then one of the primary goals of economic policy should be to assist in this pursuit. Not only is happiness affected by these policies but even institutional conditions such as the quality of governance and size of social capital affect individual level well being, arguably more so than government policy and economic expansion (John Helliwell 2001). For instance welfare policy may serve to provide a destitute individual with an elevated income level, however that may fail to resolve issues with unhappiness if that discontent is related to being unemployed. An enhanced understanding of the economics of happiness could also serve to understand aspects of human behaviour that's related to the economy. For instance the effects of happiness on consumption behaviour (Barbara Kahn and Isen 1993), work behaviour (Michelle Iaffaldano and Paul Muchinsky, 1985), and investment behaviour. Better insight into individual level well being would also lead to improvements into the microeconomic theory of utility measurement. Especially in respects to predictions of future utilities as well as the consistency of self predicted and remembered utilities (Kahnmen et. al 1997).

## Literature Review

Early economists such as Mill and Smith have always defined happiness in terms of utility which was broadly outlined in terms of material consumption and rational decisions in accordance to a monetary budget line. It was not until very recently that this view was challenged. Branching off from psychology, recent economists have begun to look into the concept of a more subjective measure of well being. Though ordinal utility has been regarded by standard economic theory as "unscientific" due to a lack of observable objectivity; cardinal utility as measured in tangible goods and services can likewise be regarded as unrepresentative of personal well being. Traditional work into utility theory has used an axiomatic approach, employing the techniques of revealed preference and cardinal utility to measure individual level utility and social welfare (Daniel Slesnick 1998). However over the years there has been a vast amount of literature questioning the validity of such an approach. Namely it is dubious that general utility can be measured objectively when utility in itself is such a subjective matter (Richard Thaler 1992). Furthermore although materialistic wants is one side of personal well being, quantifying the utility derived from it at an individual level is unfeasible.

The other sides to well being could be a vast array of factors including intrinsic beliefs, outside circumstances, and the situation of others. The subjective approach to utility is able to capture the interactions of a multitude of factors each playing a role in personal well being. Although it may be considered questionable whether people are capable of interpersonal comparability and producing unbiased answers, there has been research that has shown the contrary (Kahneman 1999). Likewise it has been found that people who claim to be happy are more likely to be considered so by friends and family members (Sandvik et al. 1993). Also where as the axiomatic approach only focuses on outcome utility, the subjective approach also encompasses procedural utility which is arguably more eminent ( Frey et al. 2004). It is altercated that individuals care at least as much about the process of obtaining utility as the acquisition of that utility. For instance although individuals may care a great deal about wealth, they are also concerned with how that wealth is accumulated.

Recent research into national happiness by Helliwell found evidence in the World Values Survey for the decreasing marginal utility of income, namely that moving up deciles in the distribution of family income increases happiness by lessening amounts (Helliwell, 2001). Other research (Di Tella, Macculloh, and Oswald, 2000) used panel data from the Eurobaronmeter to measure the effects of national economic indicators such as unemployment and inflation on predicted happiness across nations. The predicted happiness levels they constructed were based on residual values obtained from a microeconomic regression. Following in the footsteps of these researchers this paper will likewise construct a measure of national happiness, but the construction will be based on the intercept values as opposed to residual values. Also whereas previous work has been focused on cross country research, this paper focuses on how happiness has varied across time in a single country. This will provide better insight into the relationship between the independent variables on happiness as cross country differences will not be a factor.

## **Data and Methodology**

The self reported happiness levels are obtained from the General Social Survey (GSS) of the United States. The GSS, originating in 1972 and continuing to present day, contains a multitude of demographic, behavioural, and attitudinal questions and serves as a valid and scientific source for tracking the structure and development of American society (norc.uchicago.edu*)*. This pooled time series cross sectional data set contains information on individuals over a span of over 38 years. The survey question of particular interest for the purpose of this paper is: "Taken all together, how would you say things are these days – would you say that you are very happy, pretty happy, or not so happy?" This question was asked for 24 out of the 38 years, and included answers from 32701 individuals. Although happiness may not be the only possible measure of personal well being, and some would suggest life satisfaction as a better one; Blanchflower and Oswald (2000) have shown using a data set for Great Britain containing both types of questions that estimated life satisfaction and happiness equations have almost identical structures and near mirror results. Macroeconomic data was collected from the official World Bank website and political data from the US Census Bureau. Time specific dummy variables were constructed with reference to america.gov (the official US historical time line as constructed by the government).

A two step methodology will be employed in this paper; first individual perceived happiness will be regressed on socio-economic and demographic characteristics for the purpose of constructing a measure of average national happiness. The following step will involve the regression of these averages on national economic indicators to infer the role of the economy on national happiness. A vast majority of literature on the economics of happiness focus on deriving residuals from this micro-econometric regression to calculate the proportion of happiness not related to individual characteristics. These residual values are then regressed upon macroeconomic factors to partial out the effects of individual/socio-demographic factors (Di Tella, Macculloh, and Oswald, 2000). We believe that such a method is redundant and inefficient because simply adding the macroeconomic factors into the micro regression can partial out the personal variables. (The results of this regression involving both micro and macro factors can be found in Appendix C). Needless to say using residual values in a secondary regression can be tricky due to the complex nature of residuals. It's difficult to conjecture the relevant proportion of the residual due to national economic indicators, and in the event that this proportion is trivial, the results will lack significance and meaning. It's for these reasons that it is believed that using the intercept values from the year to year micro-econometric regression will instead yield more meaningful results.

The intercepts obtained from this regression for each individual year can be interpreted as the proportion of happiness that cannot be accounted for by the weighted average of microeconomic and demographic factors. The argument being as follows, for each year $\hat{\beta}_0$ takes the following form:

$$\bar{y} - \sum_{i=1}^{n} \hat{\beta}_i \bar{x}_i = \hat{\beta}_0$$

Where $\bar{y}$ represents the average happiness level defined as the sum of all individual level reported measures of happiness divided by total observations for that particular year. It is worthwhile to note that $\bar{y}$ encompasses all factors, observable or not, that lead individuals to report a given happiness level. By deducting the weighted average effect of all socio-economic and demographic variables within the micro-econometric regression, it is possible to take out these factors that are affecting the average happiness level. What remains is the average national happiness value for a particular year less the socio-demographic factors, $\hat{\beta}_0$. We speculate that these remaining yearly values of mean happiness can be partially explained by movements in the economy that are captured in macroeconomic variables such as inflation, GDP per capita, and the unemployment rate. Though it is possible that these economic indicators could also correlate with time specific events such as national disasters or technological breakthroughs creating an endogeneity problem. For this reason $\hat{\beta}_0$ will be regressed on all the preceding national economic variables and specific event dummy variables to allow for more robust measures of economic performance on the aggregated mean happiness. The precise regression of interest took the form:

$Ħ_t = \beta_0 + \beta_1 \text{Unemployment}_t + \beta_2 \text{Inflation}_t + \beta_3 \text{GDP}_t + \beta_4 \text{d\_presidency} + \beta_5 \text{ d\_disaster} + \beta_6 \text{d\_tech} + v_t$

Where the y-variable, $Ħ$ is the mean yearly happiness $\hat{\beta}_0$ obtained from the micro-econometric regression. The parameters of interest are unemployment, inflation and GDP per capita. A hypothesis test will be exercised to determine respectively whether $\beta_1$, $\beta_2$, and $\beta_3$ are statistically significant.

## **Discussion of Explanatory Variables**

The first stage of our regression involves regressing socioeconomic and demographic characteristics on individual reported levels of happiness. The variables that were chosen for specific purposes are health, marital status, employment status, family income, number of children, ethnicity, and age.

The casual interaction between health and happiness has been well documented for many decades now. It is an established fact that health affects one's happiness and vice-versa. Worsening health is more likely to cause a pessimistic view of life in general as one faces more restrictions on day to day activities due to either physical or mental inadequacy. Whereas excellent health will bring about a more cheerful stand point in life and cause greater contentment in daily life (Borghesi and Vercelli, 2008). Marital Status could likewise affect happiness in both directions depending on whether one is married. Through survey evidence Easterlin has shown that marriage in fact has a lasting positive effect on happiness and that divorce or the loss of a loved one is similarly significantly negative. The interactions between employment and happiness are also of a complex nature. Unemployment would certainly cause income loss for an individual which would inhibit consumption and henceforth lead to a reduction in objective utility. It could perceivably also affect the individual at an emotional level in the form of loss of self confidence or feelings of self inadequacy. The effects of income on happiness are a near parallel to that of the effects of unemployment. Lower income earners may feel inadequate and less confident compared to wealthier colleagues and may be disappointed with their current level of consumption. Although it's hard to establish a causality relationship between personal employment status/income and the individual happiness level, it is unambiguous that those two variables may have a strong correlation. Children are an important aspect of any household and can be the cause of joy or misery. On the one hand they serve as a source of fulfillment and signify a successful family. However for the financially struggling family, children can act as a burden that's too heavy to carry. Therefore it is hypothesized that the number of children will increase happiness until a certain number has been reached conditional on income. Ethnicity is an innate characteristic of an individual that likely influences their well-being status. There could be a genetic component to happiness that varies across differing races; it's more plausible that ethnicity affects individual happiness in relation to those around the individual, especially for immigrants, who are more likely to feel depressed in a country where most of the inhabitants are of a different race. Similarly discrimination due to ethnicity could cause a detrimental effect on the well being of the individual being discriminated against. Finally, age can to be an important personal factor for happiness due to the "U Bend Of Life Hypothesis": which states that people start off their adult life relatively happy but that levels begin on a gradual decline due to work pressures and biological changes until they reach their mid-life crisis, after which happiness rises with age. Empirical evidence for this matter has been discovered through psychological experiments by Arthur Stone.

The macroeconomic regression aims to capture the interactions between the economic indicators inflation, unemployment, and GDP per capita on average national happiness. However due to the unified movement of economic performance and national happiness; a variable contained in the error term that affects national happiness is likewise apt to affect the economy, creating a highly probable endogeneity problem. Since the dependant variable reflects a highly complex and subjective measure there exists a multitude of factors that could potentially shift it. In addition a majority of these factors are likely to be time or event specific due to the nature of the dependant variable, which was derived on a year by year basis. The introduction of time and event specific dummy variables was proposed for the purpose of resolving this issue. The party dummy reflects which party was in power, and is significant in the sense that parties often create policies based on their party's historical perspectives and such policies usually affect everyone in the nation. The disaster dummy keeps track of years where major tornados, earthquakes, or snow storm occurred, costing at least $50 million in damages. Lastly the tech variable captures monumental advances in technology (such as the development of the internet), medicine (such as the small pox vaccine), and legislation (such as the legalization of abortion). Although these dummies cannot completely reflect all time specific events that may affect the national happiness level, it's presumed that they will capture the most influential ones.

# Econometric analysis

*OLS analysis on microeconometric regression*

Stage one of our micro regression required regressing individual level happiness on the key socio-demographic factors reported in the variable discussion section above. This involved a pooled time series cross sectional regression of individual level data in order to establish an informed inference as to the relevance and significance of each variable in influencing happiness. The significant variables were then input into a second stage micro regression where each year was individually regressed to estimate the mean yearly happiness.

Since the answers to happiness surveys are ordinal instead of cardinal it is argued that they're best estimated using ordered probit/logit model; however the sign and significance of estimated coefficients produced from both types of regressions based on the same equation are remarkably similar (see appendix c). Also while the exact effects of these independent variables on happiness may not be directly observable, happiness researchers have found that OLS coefficients are an effective means to assigning weight to them (Carol Graham, 2005). Note that the main use of measuring happiness is not to compare levels in an absolute sense but rather to seek to identify the determinants of happiness. For that purpose, it is neither necessary to assume that reported subjective well-being has cardinal measurability nor that it is interpersonally comparable. Our pool time series cross sectional regression model took the following form: Happy= $\alpha_0 + \alpha_1 age + \alpha_2 educ + \alpha_3 childs + \alpha_4 d\_male + \alpha_5 d\_exc + \alpha_6 d\_good + \alpha_7 d\_poor + \alpha_8 d\_married + \alpha_9 d\_dws + \alpha_{10} d\_work + \alpha_{11} d\_unemp + \alpha_{12} d\_white + \alpha_{13} d\_black + \alpha_{14} d\_income2 \ldots + \alpha_{18} d\_income6 + \alpha_{19} d\_74 + \ldots + \alpha_{42} d\_10 + \mu$

Dummy variables for sex, health, marital status, race and income were used to better compare and understand the effects of happiness on different individuals. (The major variable description is in Appendix A, while the statistic summarization of variables is in Appendix B). The time dummies are added to capture any time/year- specific events that may affect the average happiness level, including major economic recessions, technology boosts, or policy adjustments.

The results for this regression can be found in Appendix C. It was found that the coefficient on age was highly significant and positive and that an increase of age by 1 year will approximately increase happiness by 0.2% [1] on average. (Calculated by dividing the estimated coefficient by the sample average of happiness, approximately equals to 2). Though this may seem to contradict the "U bend of life hypothesis", it actually provides concrete support. The average age for our sample taken as a whole is approximately 45. This is exactly around the bottom of the U bend, the time for the mid-life crisis, and hence increasing returns for age in terms of happiness. The results also imply that children significantly reduce happiness; increasing the number of children by one will decrease the happiness level on average by 0.24%, which should not come as a surprise given that the average income category is 4.3, suggesting most individuals fall into the $15,000-$20,000 yearly income range. These individuals are arguably financially unfit to raise children and hence suffer because of instead of enjoy the company of their progeny. Taking people of fair health to be the comparison/base group, it can be seen that improving health has the most impact on individual happiness. People of excellent health are predicted to be approximately 20% happier than those with fair health, and people with poor health are 8.25% unhappier than the base group. Income follows a fairly straightforward pattern; increasing happiness by larger amounts as a higher level of income is obtained, reaching a climax of 3.5% at the highest income bracket. This is consist with the results of Blanchflower and Oswald (2000) who found that richer people will on average report higher levels of personal well being. Marriage had the second largest impact on happiness and married individuals were on average 10% happier than those who have never been married, individuals who were in the DWS category are 5% less happier than the base group (people never get married). The negative sign of coefficient on the work dummy implies people who are not in the labour force are happier than those who work. This makes intuitive sense since a good fraction of those not in the labour force choose not to be (e.g. retired or in school), and are not subject to work pressure or job dissatisfaction. The coefficients on all other variables are consistent with our predictions and don't merit an explanation. More impressively, the individuals who are unemployed are on average 9% less happy than the group that's not in the labour force, and 6.5% unhappier than the working group.

---

[1] The estimated coefficient on variable age equals to .0038779, to better show the significance of how age affects happiness level that's out of a level three, we divide this value by the sample mean of perceived happiness which is equals to 2.19 (approximately 2) to construct that percentage value. This method of calculation is same for every other regressor.

*Discussion of Macro-economic Regression*

The following step in our two stage regression was to then regress the mean yearly happiness for each year on its' corresponding macroeconomic variables, including GDP per capita, unemployment inflation, and time-specific events. As discussed above this will allow for an interpretation of the role of national economic indicators on the national average happiness level absent the weighted average of personal explanatory variables.

Due to the nature of GDP per capita being highly correlated with time, even after linearly detrending GDP per capita the first order autocorrelation being about 0.83, suggesting that this time series may exhibit unit root behaviour (the rho_hat is however still below 0.9), and may raise serious questions about the use of usual OLS t statistics, due to the violation of CLM assumptions of the OLS estimators. Alternatively, it's believed that this variable will on average only affect individual happiness on a comparative basis; hence people measure their consumption/income well being with respects to a frame of reference (i.e. a previous time period). Therefore it'll be possible to measure how much extra income/consumption the average individual will have with comparison to last year. Henceforth taking the difference of GDP per capita between two consecutive years and constructing a raw change in GDP variable (GDPD) will provide a better understanding of its' relationship with national happiness. Under this method the time trend of GDPD is no longer as significant at the 5% significance level, and the first order autocorrelation shrinking to 0.21. After running the regression using GDPD instead of GDP per capita, the null hypothesis of the significance of this parameter is not rejected at the 5% significance level, indicating no noteworthy connection between the two variables statistically. Although the economic significance of the parameter may be greater; an increasing GDP per capita could have implications for personal wellbeing through the channel of an increasing inflation rate. Since most individuals care only for their own level of income, the increase of GDP per capita should only have an effect at the personal level to the degree that private income is affected. Alternatively, an increasing GDP per capita does not assure that all people within the nation will be wealthier; an unequal distribution of national wealth may leave the majority of common workers no better off than before the GDP increase. Should these common workers realize that a small corporate elite are gormandizing most of the increase in national wealth, this could exacerbate feelings of wellbeing. Hence it can be inferred that the effect of GDP on national happiness is an indirect one through inflation and comparison amongst individuals.

Upon testing for time trends it was discovered that inflation is in fact correlated with time, however this issue was resolved by inserting a time trend t into the main regression. The first order autocorrelation for this time series is not significant, although there was a strong relationship between inflation and GDP per capita. All of this leads to the conjecture that inflation is responding to an increase in GDP (See Log Filer). After regressing the mean yearly happiness with indicators of economic performance it was found that inflation was relatively statistically significant ($p=0.017$). In fact a 1% increase in inflation on average would roughly result in a 3.1% reduction in national happiness (taking 1.7 to be the yearly happiness average without personal factors). The intuition behind this result is straight forward, an increase in inflation results in a decrease in purchasing power due to a rise in prices. Assuming nominal wages remain constant; this is the equivalent of a decrease in real income. Recalling that private income significantly influenced happiness in the micro-regression, it comes as no surprise that national inflation affects average national happiness by distorting the purchasing power of individuals.

Unlike GDP, the variable unemployment does not have a significant time trend, but it follows a weakly dependent time series ( rho_hat=0.54). Unemployment was found to be significantly negative at the 5% level (two-sided), having a p-value of about 0.011. As touched on in the variable discussion section, unemployment indubitably affects personal wellbeing and consequentially happiness. Therefore it comes as no surprise that an increase in the unemployed will result in lower levels of national happiness on average. Though the results imply that a 1% increase in the unemployment rate on average decreases national happiness by 5.37%, it could be argued that the true effect of unemployment is being undermined since unemployment had already been controlled for in the micro-econometric regression. Following in the footsteps of Ditella and Oswald (2000), we seek to calculate the true effect of unemployment by working out the sum of the aggregate and personal effects. The coefficient of being unemployed in the micro regression signifies the cost of unemployment for the individual who are actually jobless, hence it's a personal effect. The aggregate effect on the other hand measures the consequence of a higher national unemployment on the average citizen who can either be employed or unemployed. The rationale being that a higher rate of national unemployment induces the fear of being unemployed on the common working man and also magnifies the grief of unemployment people by indicating even a lower chance of finding a job. Hence the net effect of a one percent increase in unemployment is equal to an approximate $0.01 \times 0.1759 + 0.091 = 0.0928$ reduction in happiness. The first term represents the reduction in well being for those that are unemployed and being affected by a 1% increase in unemployment. The latter is the effect on the employed people who are being scared by an increase in the national unemployment rate. Taking the average happiness level to be about 1.7, this is the equivalent of a 5.5% reduction in national happiness. Note that the aggregate effect greatly outweighs the personal effect; this is intuitively sound since the personal effect is only relevant to a small fraction of the population whereas the aggregate effect encompasses everyone.

Due to the nature of time series data it was imperative to test for serial correlation and heteroskedasticity in order to obtain meaningful results. To detect serial correlation in the overall regression, Durbin's Alternative method was implemented. The results show that there is no statistically significant serial correlation existent in this model. This doesn't imply a lack of endogeneity but rather that statistical inference is more applicable due to higher validity of the standard errors. Using the Bruesch-Pagan test, it was possible to check for heteroskedasticity (see Appendix D). Although both of these problems don't pose an issue for the consistency or unbiasedness of the OLS estimators, they will cause inefficiency due to inaccurate measures of the standard error. This in turn would result in the OLS estimators no longer being BLUE, and the construction of confidence levels, or t /F statistics would no longer be plausible. After running the Bruesch-Pagan test, the p-value was significantly larger than 5%; therefore, the null hypothesis of homoskedasticity cannot be rejected. (Result shows in Appendix D)

## **Econometric Issues**

Due to the nature of research involving the economics of happiness, there exist many issues involving the data, the subjects in the survey as well as the survey itself, and relationships amongst variables. As survey data revolve around individual judgements, they are prone to a multitude of both systematic and non-systematic measurement errors. The capability and willingness of people to give meaningful answers to questions about their reported subjective well-being may depend on the order of questions, the wording of question, scales applied, as well as momentary mood. For instance if the questions preceding the inquisition of wellbeing invoked feelings of desolation (i.e. the passing of a relative or mental/psychological state) the respondent would be more apt to report lower levels of well being. The scale of 3 also poses a problem as it does not allow for as precise of answers as a larger scale. As previously noted our question of interest only asked for the respondent's well being within the past few weeks; however this leads room for circumstantial and immediate events to bias answers causing random measurement error. However it is maintained that such data is appropriate since it has long been tradition in utility economics to rely on the judgements of peoples directly involved. This paper introduced a vast amount of explanatory variables in an attempt to reduce the unexplained portions of individual well being, and give an accurate account of average national happiness.

Sampling selection is another aspect of concern for this study as the sample may not be a true reflection of the general population of which its' allegedly representing. As mentioned in the OLS analysis section the majority of the respondents in the survey are middle aged, low income earners, and married. The correlation of measurement errors with individual characteristics could be of particular concern when there are issues with sampling selection. As an example it is possible that age has an influence on how people react and respond to questions about their subjective well-being. However this is an example of exogenous selection and such selection based on regressors does not affect the consistency of OLS estimators. Whereas individual characteristics that occur in the unobservable error term of the microeconomic regression could potentially cause an omitted variable problem; such a problem could be remedied through the use of panel data as intrinsic errors independent of time could be eliminated, and provide for more consistent results.

The establishment and direction of causality is of particular concern in this paper due to the possible existence of simultaneity bias. At the individual level it is uncertain whether certain socio-demographic variables have an impact on happiness or if the reverse effect is true. For instance does higher levels of income make one happy or do happy individuals tend to become wealthier? At the national level it is unclear whether movements in happiness correspond with macroeconomic indicators or if causality runs in the opposite direction. It could be the case that the economy tends to perform better when the nation as a whole are happier. An increasing unemployment rate could be the result of more disgruntled individuals who are in turn more prone to lose their jobs.

Endogeneity is a prevalent concern in both the microeconometric and the macroeconomic regression, as the error term may be correlated with the explanatory variables in the regressions. For the microeconometric regression, having over 30,000 data points it was considered reasonable to add as many observable personal characteristics as possible to extract factors inside the error term. However, for the macroeconomic regression, the restriction of a smaller sample size constricts the use of numerous regressors. It is on these grounds that the event specific dummies political party, national disasters, and tech were introduced; as these variables were deemed the most crucial time specific events in shaping national well being. Despite implementing this method, endogeneity may still persist, and the only permanent solution is the addition of an instrument variable or a good proxy for the unobserved factors. It should be noted however that present day research into the economics of happiness has not discovered any applicable proxies or instruments due to the complex nature of happiness (Praag, 2007).

## **Conclusion**

The aim of this paper was to discover correlation between national happiness for the general population of the United States and economic performance using individual level data from the General Social Survey and national data from the World Bank. The average happiness level for each year not attributed to individual characteristics of the respondents was obtained from a microeconomic regression of socio-demographic variables. These mean happiness levels were then regressed against indicators of national economic performance. The results imply that much of individual well being can be predicted using observable and measurable characteristics. Also there is a significantly negative connection between national happiness and unemployment and inflation. This is consistent with previous research into the economics of happiness that used residual values from a microeconomic regression of happiness on national economic indicators. Although national wealth measured in terms of GDP per capita had no statistically significant correlation with the mean national happiness, individual level income was highly significant for individual wellbeing. This in turn further highlights and adds weight to the debate regarding whether wealthier nations are indeed happier, and suggests that national economic performance only affects individual well being to the extent that the individual is directly affected. These results have serious implications for further research into welfare economics. As mentioned in the previous section research still needs to be done on the effects of happiness on behaviour to resolve the simultaneity bias problem. So far only a theoretical investigation into this issue has been conducted (Benjamin Hermalin And Alice Isen 1999). Further research into the economics of happiness on less developed countries has yet to be undertaken; such research could provide valuable insights as to how happiness varies between OECD and developing countries. The implications of a deeper understanding of how economic performance correlates with individual and national well being is imperative for evaluating the effects of government expenditure and policy, and will lead to a stronger political and economic system.

**Appendix A Major variable Definition**

| Variable | Storage type | Variable label/definition |
|---|---|---|
| happy | byte %8.0g | Respondent's perceived happiness level, 1 for not too happy, 2 for fairly happy and 3 for very happy |
| Age | byte %8.0g | Age of respondent |
| sex | byte %8.0g | Respondent's sex, 0 is female (base group), 1 is male |
| race | byte %8.0g | Respondent's race, 1 is white, 2 is black , 3 is other(base group) |
| educ | byte %8.0g | Highest year of school completed by respondent |
| Marital | byte %8.0g | Respondent's marital status, 0 is divorced/widowed/separated, 1 is never been married (base group) , 2 is married |
| Health | byte %8.0g | Respondent's condition of health, 1 is poor health, 2 is fair health(base group) , 3 is good health and 4 is excellent health |
| Work status | byte %8.0g | Respondent's labour force status, 0 is not in the labour force (base group), 1 is unemployed and 2 is working either full time or part time |
| Family income | byte %8.0g | Total family income, 1 is income from $0 to $4999 (base group), 2 is income from $5000 to $9999, 3 is income from $10000 to $14999, 4 is income from $15000 to $19999, 5 is income from $20000 to 24999, 6 is a income of $25000 or more |
| childs | byte %8.0g | Number of children in respondent's family |
| Mean yearly happiness (B0_hat) | float %9.0g | National mean happiness that's not determined by the weighted average of individual socio-economic and demographic characteristic, but only determined by time specific national factors or events. |
| unemployment | float %9.0g | National/ aggregated unemployment rate in each corresponding year |
| inflation | float %9.0g | National/ aggregated unemployment rate in each corresponding year |
| GDP_capita | float %9.0g | Nominal GDP per capita value in each corresponding year |
| Party | float %9.0g | The political party that the US president represent for in each corresponding year, 1 is democratic party and 0 is republican party(base group) |
| disaster | float %9.0g | 1 if earthquakes, hurricanes, and snow storms costing at least 50 million dollars in damages,0 otherwise(base group) |
| tech | float %9.0g | 1 indicates a technological/medical breakthrough or drastic legislation change. Including publically available internet, discovery of the small pox vaccine and legalization of abortion by the supreme |

| | | court. 0 otherwise(base group) |

**Appendix B: Summary of Statistic including dummy variables**

| Variable | Obs | Mean | Std.Deviation | Min | Max |
| --- | --- | --- | --- | --- | --- |
| happy | 32701 | 2.190973 | .6372335 | 1 | 3 |
| age | 32701 | 44.8727 | 17.07379 | 18 | 89 |
| sex | 32701 | .4511789 | .4976184 | 0 | 1 |
| Race | 32701 | 1.222073 | .5084999 | 1 | 3 |
| educ | 32701 | 12.74062 | 3.154512 | 0 | 20 |
| marital | 32701 | 1.300021 | .8426965 | 0 | 2 |
| health | 32701 | 1.978869 | .8429669 | 1 | 4 |
| Work status | 32701 | 1.28201 | .9313977 | 0 | 2 |
| Family Income | 32701 | 4.333537 | 1.807689 | 1 | 6 |
| # of children | 32701 | 1.951959 | 1.8011 | 0 | 8 |
| mean happiness | 24 | 1.688297 | .214912 | 1.349552 | 2.243827 |
| unemployment | 24 | 6.370833 | 1.493021 | 4.0 | 9.7 |
| Inflation | 24 | 4.860837 | 3.077374 | 1.552279 | 13.50937 |
| GDP per capita | 24 | 23800.96 | 12949.93 | 6461.736 | 47208.54 |
| Party | 24 | .2916667 | .4643056 | 0 | 1 |
| disaster | 24 | .25 | .4423259 | 0 | 1 |
| Tech | 24 | .125 | .337832 | 0 | 1 |

**Appendix C Micro-econometric regression**

Dependent variable: happy, individual level perceived happiness

|  | (1)OLS w/o time dummies | (2) OLS with time dummies | (3)oprobit with time dummies | (4)OLS pooled time series regression with micro and macro factors w/o time dummies |
|---|---|---|---|---|
| age | .003716* | .0038779* | .0075489* | .003876 * |
|  | (.0002425) | (.0002436) | (.0004789) | (.0002436) |
| childs | -.0046065 | -.0048119 * | -.00916 | -.0048169* |
|  | (.0021371) | (.0021369) | (.0041815) | (.0021358) |
| educ | .0029397* | .0033947 * | .006652 * | .0033559 * |
|  | (.0012167) | (.0012195) | (.0023893) | (.001218) |
| d_male | -.0423471* | -.0440711* | -.0864579* | -.0439281* |
|  | (.0069038) | (.0069077) | (.01353) | (.006903) |
| d_excellent | .3892919* | .3834262* | .7450451* | .3836559 * |
|  | (.0101429) | (.0101684) | (.020115) | (.0101705) |
| d_good | .1876078* | .1850629 * | .3525973* | .18549* |
|  | (.0093267) | (.0093232) | (.0181566) | (.0093249) |
| d_poor | -.1659433* | -.1651276 * | -.3151924 * | -.1645805* |
|  | (.0163003) | (.0162923) | (.031701) | (.0162896) |
| d_married | .2042828* | .1902662 * | .3678179 * | .1907222* |
|  | (.0099313) | (.0102045) | (.0199444) | (.0101733) |
| d_DWS | -.0910803* | -.0941542 | -.1819048 * | -.0942496* |
|  | (.0114791) | (.0114908) | (.0223515) | (.0114851) |
| d_work | -.0503782* | -.0515756 * | -.1022616* | -.0516208* |
|  | (.0083862) | (.0083844) | (.0164629) | (.0083825) |
| d_unemp | -.1759622* | -.1742181 * | -.3371465* | -.1733716 * |
|  | (.0158685) | (.0158701) | (.0309801) | (.0158678) |
| d_income2 | .0018328 | .001804 | .0028937 | .0017402 |
| | (.0142641) | ( .0142757) | (.0277501) | (.0142682) |
| d_income3 | .0217336 | .029907 | .0568693 | .0300735 |
| | (.0144173) | (.0144926) | ( .0282106) | (.0144702) |
| d_income4 | .0209541 | .0327179 | .0629357 | .0318749 |
| | (.015475) | (.0156434) | (.0304806) | (.0155704) |
| d_income5 | .0335043 | .0518195 * | .0996065* | .0501889* |
| | (.0158044) | ( .016163) | (.0315253) | (.0160184) |
| d_income6 | .0711488* | .1036277* | .2014263* | .1011117 * |
| | (.0134098) | ( .0145367) | (.0283283) | (.01431) |
| d_white | .0324965 | .019174 | .0371619 | .0178356 |
| | (.0163431) | ( .0165396) | (.0322556) | (.0165264) |
| d_black | -.0775754* | -.0848869* | -.1625801 | -.0866461* |
| | (.0182679) | (.0184019) | (.0358375) | (.0183527) |
| d_74 | - | .0116159 | .022478 | - |
| | | (.0226667) | ( .0446107) | |
| d_75 | - | -.0300467 | -.0608161 | - |
| | | ( .0224562) | ( .044046) | |
| d_76 | - | -.0038539 | -.0087798 | - |
| | | (.0225121) | ( .0442292) | |
| d_77 | - | .0082696 | .0153485 | - |
| | | (.0225819) | ( .0443753) | |
| d_80 | - | -.0276184 | -.0556055 | - |
| | | (.0228247) | ( .0447647) | |
| d_82 | - | -.0333653 | -.0650624 | - |
| | | ( .0217972) | (.0427492) | |
| d_84 | - | -.0145281 | -.029497 | - |
| | | (.0231707) | (.0454817) | |
| d_85 | - | -.0726748 * | -.1433293* | |

| | | ( .0227464) | (.0445817) | |
|---|---|---|---|---|
| d _87 | - | -.0549154 * (.0221511) | -.1083856* (.0434187) | - |
| d _88 | - | .016083 ( .0258809) | .029923 (.0508431) | - |
| d _89 | - | -.0458689 (.0256543) | -.0918827 . (0503044) | - |
| d _90 | - | -.0238638 (.0264729) | -.0475511 (.0519744) | - |
| d _91 | - | -.0507245 (.0259931) | -.1015342 (.050892) | - |
| d _93 | - | -.0618447* ( .0252453) | -.120847 * (.0495093) | - |
| d _94 | - | -.0980349* (.022008) | -.1912681 * (.0431489) | - |
| d _96 | - | -.0573505 * (.0212876) | -.1123058 * ( .0417458) | - |
| d _98 | - | -.0544503 * (.0208144) | -.1057289* (.040856) | - |
| d _00 | - | -.0299065 (.0216283) | -.0582165 ( .042464) | - |
| d _02 | - | -.041965 ( .0268814) | -.0820195 ( .0527097) | - |
| d _04 | - | .0468132 (.0273151) | -.0920738 ( .0535732) | - |
| d _06 | - | -.056427 * (.0225336) | -.1107358* (.044204) | - |
| d _08 | - | -.1111186 * | -.2174299 * | |

|  |  | (.0244358) | (.0478803) |  |
|---|---|---|---|---|
| d_10 | - | -.1264708 * (.0246132) | -.2460712* (.0481356) | - |
| unemp2 | - | - | - | -.0088134* (.0022813) |
| Infl | - | - | - | .0018459 (.0015767) |
| GDP_capita | - | - | - | -1.99e-06* (3.94e-07) |
| Constant/ Cut for oprobit | 1.712237* (.0264562) | 1.748234* (.0307054) | -.4087738(cut 1) 1.383874(cut 2) | 1.801869 ( .0341763) |
| Adjusted R squared | 0.1369 | 0.1388 |  | 0.1383 |
| Pseudo R squared Likelihood ratio | - | - | 0.0786 4885.25 |  |

→ The symbol "*" applies for that the estimated coefficient is statistically significant under a 5% significance level (double sided)

→For the aggregate regression of combined micro/macro economic variables different results were obtained as opposed to the 2 stage regression which was the focus of this paper. Though unemployment is still significantly negative, inflation is on the contrary no longer so. This could be the result of the respondent's income level explaining part of the significance of inflation. As mentioned in the paper inflation only affects individuals in the sense that individual level income changes. The difference in results can mainly be attributed to diverse dependant variables. Whereas the dependant variable in this regression is individual level happiness, the dependant variable in the 2 stage regression was average yearly happiness. Hence the results imply that individual happiness may vary from national happiness.

**Appendix D Macroeconomic regression result**

Dependent variable: B0_hat, the mean yearly happiness

|  | (1) OLS regression of time series absent time/event specific dummies | (2) OLS regression of time series data with time/event specific dummies |
|---|---|---|
| Unemp | -.0679655 | -.0913312 * |
|  | ( .0326956) | (.0317144) |
| Infl | -.0450894 | -.0526246 * |
|  | ( .0200995) | ( .0195535) |
| GDPD | -.000061 | -.0000541 |
|  | ( .0000451) | (.0000425) |
| t | -.0138405 | -.0164794 |
|  | (.0100671) | (.0092784) |
| Party | - | -.2391785* |
|  |  | (.0935348) |
| Disaster | - | -.058875 |
|  |  | ( .0896972) |
| Tech | - | .1977392 |
|  |  | (.1587999) |
| Constant | 2.627815* | 2.907495* |
|  | (.3177976) | (.3160263) |
| Number of observation | 23 | 23 |
| Adjusted R-squared | 0.2599 | 0.4131 |
| Serial correlation rho_hat | .23367 | -.4465528 |
|  | (.2612913) | (.3348276) |
| BP test F statistics | $F(4, 18) = 1.47$ | $F(7, 15) = 0.63$ |
|  | Prob > F = 0.2539 | Prob > F = 0.7212 |